\documentclass{aa}

\usepackage{acronym}		%
\usepackage{glossaries}		%
\usepackage{booktabs}

\usepackage{fontawesome}

\usepackage{hyperref}
\usepackage{color}

\usepackage{graphicx}
\usepackage{txfonts}
\usepackage{lipsum}
\usepackage{subcaption}         %
\usepackage{lscape}             %
\usepackage{placeins}           %
                                
\usepackage{braket}
\usepackage{bm}

\hypersetup{hidelinks}

\newcommand*\diff{\mathop{}\!\mathrm{d}}
\newcommand{\inner}[2]{\left\langle #1, #2 \right\rangle}

\usepackage{amssymb}
\usepackage{pdfrender}
\newcommand*{\boldcheckmark}{%
  \textpdfrender{
    TextRenderingMode=FillStroke,
    LineWidth=.5pt, %
  }{\checkmark}%
}

\newcommand{\norm}[1]{\left\lVert#1\right\rVert}

\newcommand{\fib}{fibre} %

\newacronym{USyd}{USyd}{the University of Sydney}
\newacronym{ANU}{ANU}{Australia National University}
\newacronym{ESO}{ESO}{European Southern Observatory}

\newacronym{NSF}{NSF}{National Science Foundation}
\newacronym{LIEF}{LIEF}{Linkage Infrastructure, Equipment and Facilities}

\newacronym{VLTI}{VLTI}{Very Large Telescope Interferometer}
\newacronym{ATs}{ATs}{auxiliary telescopes}
\newacronym{UTs}{UTs}{unit telescopes}
\newacronym{ELT}{ELT}{Extremely Large Telescope}
\newacronym{JWST}{JWST}{James Webb Space Telescope}
\newacronym{DCT}{LDT}{Lowell Discovery Telescope}
\newacronym{GMT}{GMT}{Giant Magellan Telescope}
\newacronym{LBT}{LBT}{Large Binocular Telescope}

\newacronym{STS}{STS}{six telescope simulator}
\newacronym{GPAO}{GPAO}{GRAVITY+ adaptive optics}

\newacronym{WFS}{WFS}{wavefront sensor}
\newacronym{ADC}{ADC}{atmospheric dispersion corrector}
\newacronym{LDC}{LDC}{longitudinal dispersion corrector}
\newacronym{AO}{AO}{adaptive optics}
\newacronym{SNR}{SNR}{signal-to-noise ratio}
\newacronym{ADU}{ADU}{arbitrary data units}
\newacronym{RMS}{RMS}{root-mean-square}
\newacronym{PIAA}{PIAA}{phase-induced amplitude apodisation}
\newacronym{ZWFS}{ZWFS}{Zernike wavefront sensor}
\newacronym{GPU}{GPU}{graphics processing unit}
\newacronym{PAWS}{PAWS}{piston-adapted wavefront sensor}
\newacronym{MKID}{MKID}{microwave kinetic inductance detector}

\newacronym{OAP}{OAP}{off-axis paraboloid}
\newacronym{DM}{DM}{deformable mirror}
\newacronym{MEMS}{MEMS}{micro-electromechanical system}
\newacronym{IR}{IR}{infrared}
\newacronym{SLM}{SLM}{spatial light modulator}
\newacronym{OPD}{OPD}{optical path difference}

\newacronym{SMF}{SMF}{single-mode \fib{}}
\newacronym{MCF}{MCF}{multicore \fib{}}
\newacronym{MMF}{MMF}{multimode \fib{}}
\newacronym{SLD}{SLD}{superluminescent diode}

\newacronym{CAD}{CAD}{computer aided design}
\newacronym{GUI}{GUI}{graphical user interface}
\newacronym{BFGS}{BFGS}{Broyden–Fletcher–Goldfarb–Shanno}

\begin{document}

   \title{Differentiable design of the PIAA-ZWFS: a flexible wavefront sensor that approaches the fundamental limit}

   \subtitle{}

   \author{A. K. Taras\inst{1}\fnmsep\thanks{Corresponding author: taras@strw.leidenuniv.nl}
        \and S. Y. Haffert\inst{1}
        \and L. Desdoigts\inst{1}
        }

   \institute{\inst{1}Leiden Observatory, Leiden University, PO Box 9513, 2300 RA Leiden, The Netherlands
   }

   \date{}

  \abstract
   {Extreme adaptive optics is necessary for high-contrast astronomy at scales of the habitable zone of nearby systems. }
   {We seek to evaluate wavefront sensors that approach fundamental limits of wavefront sensing, enabling adaptive optics systems to run faster or on fainter targets.} 
   {We present the phase-induced amplitude apodisation Zernike wavefront sensor (PIAA-ZWFS): an adaptation of the conventional Zernike wavefront sensor (ZWFS) that leverages lossless apodisation of the pupil to concentrate the starlight in the focal plane.  
   We optimise and evaluate the sensor with a differentiable modelling framework, drawing on concepts from Bayesian experimental design to minimise the variance of a maximum-likelihood estimator that uses the system in the high-Strehl regime.}
   {Our architecture shows state-of-the-art performance in simulation for different apertures, bandwidths, photon fluxes and source sizes, closing the gap to the fundamental limit in variance by a factor 10 (2.5) compared to the conventional ZWFS (optimised ZWFS) in a typical photon-limited case. For extended sources, we show that even an ideal point source sensor rapidly becomes sub-optimal, and our system outperforms it for stellar diameters larger than 0.8$\lambda/D$ (where $\lambda$ is the wavelength and $D$ is the telescope diameter). We verify that these gains do not compromise dynamic range for either linear or nonlinear reconstructors. Also, we present a proof that there must be a trade-off between the information gained about amplitude and phase errors for any wavefront sensor.
   }
   {The PIAA-ZWFS is a viable wavefront sensor operating near the fundamental sensitivity limits in a wide range of simulations.}


   \maketitle
\nolinenumbers

\section{Introduction}

Extreme \gls{AO} systems are necessary for the detection and characterisation of exoplanets from ground-based telescopes \citep{guyon2018extreme}. In particular, instruments on the coming generation of \glspl{ELT} could extend the list of directly imaged exoplanets to include habitable zone planets around nearby low-mass stars \citep{kasper2021pcs, males_high-contrast_2024}. Assuming such a system can maintain diffraction-limited and background noise dominated performance, the exoplanet scientific yield of instruments with a telescope diameter of $D$ scales almost as $D^4$ \citep{males_mysterious_2021}, as opposed to other applications (e.g.~radial velocity surveys) which scale as $D^2$. 
Beyond ground-based instruments, more efficient wavefront sensors are becoming important for the next generation of high-contrast, space based missions, where picometre wavefront sensing is required \citep{steeves2020picometer}. For such applications the time needed to flatten the wavefront can be reduced by hours by improving the efficiency of the adaptive optics system. 

In order to achieve diffraction limited performance, \glspl{WFS} must be very sensitive to phase aberrations, enabling the control loop to function well at kilohertz rates. Different \glspl{WFS} are known to have different sensitivities to phase aberrations at the pupil plane~\citep{ragazzoni1999sensitivity, guyon2005limits}. The sensitivity limit given a number of photons has been known for some time \citep{paterson2008towards} but no architectures have been demonstrated in hardware that approach this limit. For the past two decades the phase contrast technique used in the \gls{ZWFS} \citep{zernike1934diffraction, n2013calibration} has been regarded as most sensitive in the photon-noise-limited regime. This architecture images starlight onto a mask that shifts the phase of the central 1.06$\lambda/D$ (where $\lambda$ is the wavelength of the light), resulting in interference that reveals the phase aberrations when the pupil plane is imaged. Previous work has shown that it is possible to optimise the phase mask structure to maximise the sensitivity in the linear \citep{chambouleyron2022optimizing} or, through joint optimisation with a neural network, nonlinear \citep{landman2022joint} regime. Beyond adaptive optics, previous work by~\cite{Desdoigts2026Differentiable} has framed imaging problems through the lens of Bayesian experimental design~\citep{lindley1956measure, rainforth2024modern}, optimising an optical system to best constrain system and scene parameters. We employ a similar approach in this work to the design of \gls{AO} systems.

In this work, we explore the phase-induced amplitude apodisation Zernike wavefront sensor (PIAA-ZWFS) for use in a second-stage adaptive optics system. This sensor was briefly introduced previously in \cite{haffert2023reaching}. It consists of two relatively simple modifications to the classical \gls{ZWFS}: the addition of \gls{PIAA} lenses to concentrate starlight on the phase mask, and the freedom to construct a multi-level phase mask at the focal plane.

Our approach optimises this new architecture to minimise the variance of an unbiased, maximum-likelihood estimator when estimating the phase at the pupil plane. This is similar to, but not the same as, previous work \citep{chambouleyron2022optimizing, landman2022joint} that optimised the information gained from phase aberrations at different spatial frequencies one at a time. Our method correctly captures possible crosstalk in the system and better reflects the expected residuals in closed loop operation; explicitly, the optimisation metric is the residual phase variance of the system. 

First, we develop the Fisher information framework of wavefront sensing, deriving previous results on sensitivity limits from this different perspective. Next, we evaluate the performance of the PIAA-ZWFS in a variety of scenarios: different apertures, stellar magnitude, bandwidth and stellar size. These results are compared to variants of the \gls{ZWFS} and an ideal wavefront sensor that saturates the information limit. We also show that each component of the PIAA-ZWFS is important in our ablation study, and that the system can still reconstruct the wavefront with the same dynamic range as existing sensors.

\begin{figure*}[t]
\centering
\includegraphics{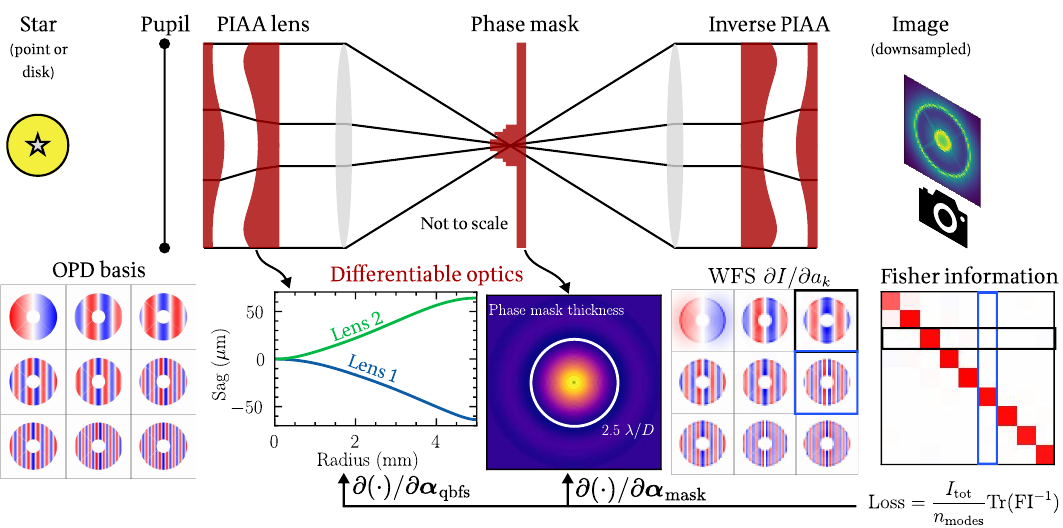}
  \caption{Architecture of the PIAA-ZWFS and our optimisation framework. We simulate starlight from a point or resolved source through a telescope pupil. This then passes through a set of PIAA lenses for lossless apodisation, is focused on a phase mask and then re-imaged in the pupil plane through an inverse PIAA lens. The final image is formed by downsampling the intensity distribution. As the simulation is differentiable, we are able to compute the derivative of the output intensity with respect to the phase coefficient (``WFS $\partial I/\partial a_k$''). All pairs of the basis set are then used to calculate the Fisher information matrix of the system. The optimiser measures the average variance of a maximum-likelihood estimator (given by the trace of the inverse of the Fisher information matrix divided by the number of modes) and is able to access derivatives of this metric with respect to optical prescriptions in the \gls{PIAA} lenses ($\bm{\alpha}_{\mathrm{qbfs}}$) and phase mask ($\bm{\alpha}_{\mathrm{mask}}$), converging on an optical design that best constrains the phase estimates. }
     \label{fig:overview}
\end{figure*}

\section{The PIAA-ZWFS}
\subsection{System Overview}
\autoref{fig:overview} depicts a pencil sketch of the PIAA-ZWFS architecture and our optimisation approach. Note that all the sub-figures show the actual solution found for the narrowband case considered in Section~\ref{sec:bw}.

The architecture contains two significant differences from a conventional \gls{ZWFS}: the pair of \gls{PIAA} lenses and more degrees of freedom in the phase mask. These modifications are well motivated by considering the \gls{ZWFS} as an interferometer, where the phase shifted beam interferes with the beam outside the phase mask. An interferometer has maximal response when amplitudes are equal and opposite phase, which requires the `mode' of the beam through the phase mask to match the beam outside the mask. For cases where the pupil is a top hat function, the optimal phase mask resembles the diffraction pattern \citep{chambouleyron2022optimizing, landman2022joint}. Optimising the phase mask without apodisation, however, requires a mask that has large extent, and hence performs worse at larger bandwidth. \cite{slepian1965analytic} first studied the problem of apodisation to suppress side lobes of the diffraction pattern. 
It is known that the Fourier transform of a function with discontinuous $m^{\text{th}}$ derivative is asymptotically a power law with exponent $-(m+1)$ \citep{bracewell1966fourier}, and hence apodisation in the pupil plane restricts the size of the phase mask necessary for mode matching significantly. 
Noting that the information gained from a wavefront sensor is measured per photon, we limit ourselves to phase-induced apodisation \citep{guyon2003phase} only, which is lossless up to reflection losses.

To fine tune the overall architecture, our optimisation approach leverages differentiable simulation. We apply a set of orthonormal \gls{OPD} probes at the pupil and compute the derivative of the wavefront sensor intensity with respect to the coefficient of each basis term. These responses are used to compute the information gained about the aberrations in the pupil plane. We optimise the system rapidly using gradient based methods, computing the derivative of this loss function with respect to lens asphere coefficients $\bm{\alpha}_\mathrm{qbfs}$ and phase mask thicknesses at different radii $\bm{\alpha}_\mathrm{mask}$. 

The PIAA-ZWFS has a relatively straightforward path towards manufacturing. \gls{PIAA} lenses have already been manufactured successfully \citep{galicher2005laboratory, jovanovic2017efficient, foster2023implementation, tonucci2026phase}, while advances in greyscale lithography \citep{grushina2019direct}, liquid crystals \citep{doelman2017patterned} or optical metasurfaces \citep{chang2018optical} allow for the creation of masks with this control of phase.

\subsection{Theory}
We begin by outlining our formulation of the wavefront sensing problem, defining the Fisher information and drawing theoretical insights about the space of possible designs.

Let the input field $E_i$ of a coherent point source entering the system be defined as
\begin{align}
    E_i(x,y) &= A(x,y) \exp(i \phi(x,y))/S_A,
\end{align}
where $(x,y)$ are the coordinates in the pupil plane, $A$ is the aperture function (1 everywhere inside the aperture and 0 outside), $S_A$ is a normalising term (such that the total power of $E_i$ is unity) and $\phi$ is the phase. In this paper (with the exception of Appendix \ref{ref:amp_and_phase_derivation}) we assume that no amplitude aberrations are present and leave this as future work. 

Drawing on previous work about the kernel formalism for wavefront sensors \citep{fauvarque2019kernel}, the output electric field $E_o$ from an optical system is defined via a linear operator $\mathcal{C}$ as
\begin{align}
    E_o (x',y') &= \iint K(x',y',x,y) E_i(x,y) \diff x \diff y =: \mathcal{C}\{E_i\},
\end{align}
where $K(x',y',x,y)$ describes the optical system and $E_o(x',y')$ is the output electric field, defined on detector coordinates $(x',y')$. Coordinates are dropped onwards for brevity. This operator is linear in $E_i$ due to the linearity of the integral. The intensity measured is just $I = N_{\mathrm{ph}}\left|E_o\right|^2$, where $N_{\mathrm{ph}}$ is the number of photons.

Let $\{W_k(x,y)\}, k\in \mathbb{N}_0$ be an orthonormal basis set with respect to the usual norm over the pupil, i.e.
\begin{align}
    \inner{W_i}{W_j} &= \iint A W_i W_j\diff x \diff y = \begin{cases} 
      1 & i=j \\
      0 & i\neq j
   \end{cases}\ \ .
\end{align}
Without loss of generality, let $W_0$ be the piston term that we do not attempt to sense.

The aberrations in the pupil plane are then expanded as a linear combination with coefficients $a_k$ to 
\begin{align}
    \phi &= \sum_{k>0} a_k W_k,
\end{align}
where the piston term is omitted as optical systems are invariant to global phase.

The wavefront sensing problem is: given measurements of the intensity distribution $I$, estimate the coefficients of phase aberrations at the pupil plane $a_k$. The Fisher information framework enables us to quantify how well constrained the coefficients $a_k$ are given a system $\mathcal{C}$ and an input field $E_i$ (assumed to be a flat wavefront for closed loop operation). 

The Fisher information of such a system under a combination of photon and read noise is
\begin{align}\label{eq:fi}
    \text{FI}_{a_k a_l} = \sum_j \frac{(\partial_{a_k}I_j)(\partial_{a_l}I_j)}{I_j+\sigma_R^2},
\end{align}
where the sum is taken over all pixels $j$ in the detector and $\sigma_R$ is the read noise of the detector. We include a proof of this and relations to previous work in Appendix~\ref{app:derivs}. While previous work~\citep{chambouleyron2023modeling} derived sensitivity in the photon or read noise cases, the above allows us to treat situations between the two, and correctly deal with cases where some pixels have more flux than others. 

\autoref{eq:fi} requires the derivative of the intensity with respect to the coefficient of the phase aberration. This can be computed efficiently by applying the chain rule (see also \autoref{eqn:chain})
\begin{align}
    \frac{\partial_{a_k}I}{N_{\mathrm{ph}}} 
    &= 2\Re\left\{E_o^* \partial_{a_k} E_o\right\} \nonumber\\ &= 2\Re\left\{E_o^* \mathcal{C}\{\partial_{a_k} E_i\right\}\} \nonumber\\ &= 2\Re\left\{E_o^* \mathcal{C}\{i W_k E_i\right\}\},
\end{align}
noting that we may interchange the derivative and $\mathcal{C}$ operators as the integration bounds do not depend on coefficients and $E_i$ is differentiable in the aberration coefficients $a_k$.
This result illustrates that the Jacobian of the intensity depends on the output electric field and the field generated by propagating $W_kE_i$ through the system. Hence, the computational cost of measuring the Fisher information of an optical model with respect to $N$ phase aberrations is $N+1$ optical propagations -- an important factor in the tractability of the optimisation framework presented. 

The information limit to the wavefront sensing problem is the bound of \autoref{eq:fi} under photon noise, which has been shown to be 4 per photon per radian \citep{paterson2008towards}. These bounds are also verified in our formalism for completeness in Appendix~\ref{app:derivs}.

We also note that Appendix~\ref{ref:amp_and_phase_derivation} presents a novel proof that the sum of the (classical or quantum) information from amplitude and phase of the electric field is bounded to 4 for all modes. Thus sensors which maximise phase information necessarily lose information about amplitude variations.

It has also been previously shown that the ideal point source wavefront sensor is one that shifts the piston mode of the system by $\pi/2$ in phase and leaves all other modes unaffected \citep{chambouleyron2024coronagraph}. We can quickly confirm this in the above framework. As the zero point of the wavefront sensor is the piston mode, the output electric field is just $E_o = iA$. As the system leaves other modes unaffected, $\mathcal{C}\{W_nE_i\} = W_nE_i = W_n$. Hence we can consider the numerator of the Fisher information (on the diagonal) in either photon or read noise cases as
\begin{align}
    4N_{\mathrm{ph}}^2(\Im(E_o^*\mathcal{C}\{W_n E_i\}))^2 &= 4N_{\mathrm{ph}}^2(\Im(iA W_n))^2 \nonumber \\ &= 4N_{\mathrm{ph}}^2(W_n)^2.
\end{align}
This expression integrates to unity per photon (as $W_n$ is part of an orthonormal basis). One can similarly show that off-diagonal elements are zero. For cases where the zero point is not a flat wavefront, the wavefront sensor can be modified with a phase transformation that first makes the desired zero point become a flat wavefront, and then the piston mode is shifted in phase as above. 

Hence an ideal point source wavefront sensor saturates both the photon and read noise bounds. While the sensitivity to either does not need to be the same in general for an arbitrary wavefront sensor, it is in the case of the ideal wavefront sensor. 

We use a (lossless and achromatic) simulation of such a device to illustrate the limits of all wavefront sensors in the numerical results in the following sections. This ensures any loss of sensitivity due to downsampling occurs equally in the ideal case as all others. Also, we use broadband simulations and hence define the phase aberration basis in \gls{OPD}, as this better reflects the nature of the atmosphere. 

Note that we have introduced this concept as the ideal point source wavefront sensor because, as we show, this wavefront sensor is far from ideal when it comes to extended sources.

For any lossless, sub-optimal wavefront sensor, however, it is not possible for the Fisher information matrix cross terms $(\text{FI}_{a_k a_l}, l\neq k)$ to vanish. The only exception is a wavefront sensor that shifts the piston mode by some phase while leaving all other modes unaffected (like the ideal wavefront sensor but with a general phase shift). We provide a proof of this in Appendix~\ref{app:offdiag}. Furthermore, performance of the system is not best reflected by this information quantity directly. Rather, the performance of an adaptive optics system is typically quantified with the Strehl ratio, which in the well corrected limit depends on the \gls{RMS} of the wavefront errors. Thus, the variance of each mode in a closed loop system better reflects the expected performance. 

For these two reasons we do not optimise the Fisher information itself, instead the loss function used is the average variance of all modes when estimated via a maximum-likelihood estimate via the trace of the covariance matrix -- the inverse of the matrix with elements $\text{FI}_{a_k a_l}$. This metric is minimised and is a standard metric in Bayesian experimental design.

\subsection{Implementation in simulation}\label{sec:implementation}

The numerical simulation is written in Python\footnote{All code can be found \href{https://github.com/ataras2/piaa-zwfs-opt}{here \faicon{github}}}, leveraging the auto-differentiation provided by \texttt{jax} \citep{bradbury2021jax}. The \texttt{dLux} package \citep{desdoigts2023differentiable} is used for modelling of optical components and propagation, and has also previously been used for optimising Fisher information of astrometric measurements \citep{Desdoigts2026Differentiable}. The \texttt{hcipy} package \citep{por2018high} is used for apertures and \gls{PIAA} initialisation.

The aberration basis set used is a Fourier basis of both sines and cosines in the $x$ axis only (exploiting the near circular symmetry of the problem), with 7 frequencies equally spaced from 0.01 to 8 cycles per pupil. The 0.01 cycles/pupil cosine mode is removed as it resembles piston and then the modes are orthonormalised over the support of the aperture. Aberrations are applied in \gls{OPD}. We find the converged designs do not change significantly when altering the frequencies and/or the number of basis aberrations. A pixel basis could be used, but would miss the low frequency modes and be computationally inefficient as it doesn't exploit the symmetry of the problem.

The \gls{PIAA} lens profiles are first initialised using a heuristic from the mode matching argument described above, but then allowed to vary during the optimisation of the sensor. The initial profiles of the \gls{PIAA} lenses are found by solving for the apodisation profile that is an Eigenmode of a spatial filter with diameter 2$\lambda_c/D$, where $\lambda_c$ is the central wavelength. The profile is then fit as a Gaussian and converted to aspherical lens parameters following the procedure outlined in~\cite{hoffnagle2000design}. We employ the $Q_{\mathrm{ bfs}}$ basis from~\cite{forbes2007shape} to parametrise the possible surfaces of the aspheric lenses. This basis was originally invented as an improvement over an even polynomial expansion which was poorly suited for design and tolerancing as typical shapes required alternating signs and large coefficients, leading to numerical instabilities. The advantages of the $Q_{\mathrm{ bfs}}$ basis for tolerancing also mean that it is a good basis for optimisation. 

The phase mask thickness is defined on a basis consisting of concentric annular rings. We use 64 equally spaced annuli over 5 resolution elements (corresponding to features sizes of $\approx5\,\mu$m for a $f/$70 beam at $1\,\mu$m), with a pixel sampling of 32 pixels per $\lambda/D$ in the focal plane. For the propagation routine we leverage the semi-analytic methods presented in~\cite{soummer2007fast}. 

We note that existing approaches like different dot size \glspl{ZWFS} are contained within the parametrisation above -- a flat surface on each \gls{PIAA} lens and a phase mask that is simply a circular, axisymmetric dot. Our results show that the degrees of freedom given are indeed able to significantly improve the performance of the system. 

Simulations are typically run with 3 wavelengths over the width of the Z band (900\,nm centre wavelength, 152\,nm bandwidth), with the exception of the bandwidth variation experiment which uses 5 wavelengths. This sampling is more than sufficient to capture wavelength dependent behaviour in the image plane with diameter $\sim2\lambda/D$. For wavelength dependence, both the \gls{PIAA} and phase mask are simulated to be made from fused silica. Further optimisation of the dispersive elements is left for future work.

\subsection{Optimisation}

The complete PIAA-ZWFS model has 82 parameters: 64 phase mask thickness values, 8 asphere coefficients per \gls{PIAA} lens and one spherical term per \gls{PIAA} lens. The distance between the lenses is degenerate and hence is not optimised. 

We employ the \gls{BFGS}~\citep{fletcher2013practical} algorithm via the \texttt{optimistix} package~\citep{rader2024optimistix}. This method directly uses the gradients available and also estimates the Hessian of the optimisation function, resulting in faster convergence and robustness to different parameter scales (which are expected in problems of this nature). To keep results comparable between different experiments and provide robust optimisation, the loss function is normalised per photon and per mode. Furthermore, the model parameters are implemented in $\mu$m, such that all are closer to unity.

All optimisations except the resolved sources were executed locally on a laptop grade \gls{GPU} (6 GB NVIDIA RTX PRO 500 Blackwell), with each optimisation taking no more than a few minutes. The resolved source optimisations were run on an NVIDIA L4 \gls{GPU} and took less than half an hour for each optimisation. 

\section{Results}

As a reminder, the loss function (average variance of estimates from a maximum-likelihood estimator) used in this work has units of ($\mu$m OPD)$^2$ photons per frame. For example, a loss value of $9\times10^{-3}$ indicates that the variance of the phase coefficient fit on a frame with $10^3$ photons will be $9\times10^{-6} \,\mu$m$^2$ \gls{OPD} per mode, or a standard deviation of 3\,nm per mode.

\subsection{The PIAA-ZWFS can saturate the information bound at high spatial frequencies}\label{sec:infosat}

\begin{figure}[h]
    \centering
    \includegraphics{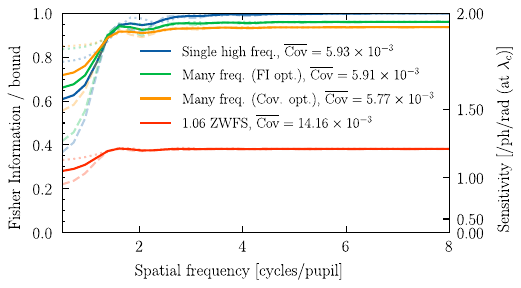}
    \caption{Verification that PIAA-ZWFS can reach the sensitivity limit for high spatial frequencies. The information gained, normalised to that of an ideal wavefront sensor as a function of the spatial frequency of the applied mode. Solid lines show the quadrature averaged information gain to a single sine (dashed) or cosine (dotted) Fourier mode. A classical ZWFS is shown in red as a baseline. Our sensors are either optimised for a single high frequency (blue), to maximise the Fisher information of multiple frequencies (green) or to minimise the covariance to multiple frequencies (orange, 7 frequencies equally spaced between 0.01 and 8 cycles per pupil). Legend text shows the average variance of an estimator of each mode over the full basis set. For comparison with other work, the right axis marks the sensitivity ($=\sqrt{\text{FI}}$ per radian at the centre wavelength).
    }\label{fig:max_diff_freqs}
\end{figure}

Here we present a verification of the results shown previously by~\cite{haffert2023reaching}, where the PIAA-ZWFS is able to saturate the wavefront sensing information limit for high frequency aberrations. We also illustrate how our approach of minimising the covariance finds a different solution. This experiment is conducted using the whole Z band on the Magellan aperture with a high pupil sampling. The results are illustrated in \autoref{fig:max_diff_freqs}, with results for a wider sweep of cases in Appendix~\ref{app:generalisation}. When optimising for a single high frequency (5 cycles/pupil, blue) the sensor reaches the fundamental limit at high frequencies, at the cost of lower information gain at low frequencies. In this case, minimising the average variance has the same solution as maximising the information gained (up to optimisation termination conditions). Optimising for the information gained at multiple frequencies (spaced as in \autoref{sec:implementation}) reduces the performance at high frequencies but improves the information gained at low frequencies. Finally, an optimisation to minimise the average variance of a range of frequencies (orange) weighs the poor performing low order terms even more, improving them at the cost of the higher frequencies yet again. The average variance values in the legend reflect (unsurprisingly) that this indeed improves the performance in this metric. Higher weights to high frequencies could drive the covariance optimised (orange) case towards the single high frequency (blue) case. While not used in this work as our metric is the variance bound of an estimator (from the Fisher information), the sensitivity of the wavefront sensor is marked on the right y-axis to allow for comparison with other publications. All designs are significantly more sensitive than the classic \gls{ZWFS} (red). 

We note one difference from the plots shown in~\cite{haffert2023reaching}: at small spatial frequencies, the Fisher information approached zero in that work, whereas it remains non-zero here. In~\cite{haffert2023reaching}, low frequency modes were not normalised and hence each Fourier mode approached the (undetectable) piston mode, while in this work the normalisation means that low frequency sine modes approach the (detectable) tip/tilt Zernike.

\subsection{Performance in different scenarios}

\subsubsection{Aperture}
\begin{figure}[h]
    \centering
    \includegraphics{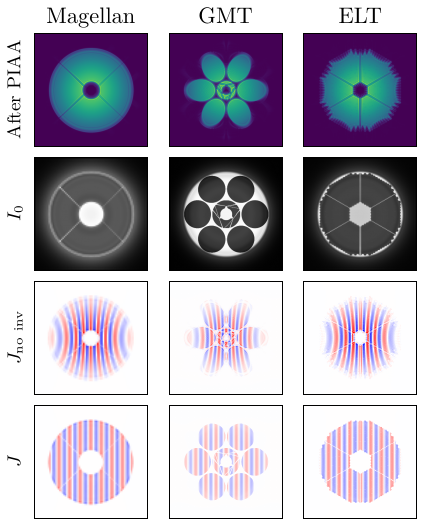}
    \caption{ PIAA-ZWFS designs and behaviour for different apertures. The apodisation profile after the first lens pair (1\textsuperscript{st} row, showing $|E|$) concentrates the light in the centre of the pupil. The reference intensity (2\textsuperscript{nd} row, now at the sampling of the detector) scatters some light outside the pupil. The intensity change for an applied Fourier mode (3\textsuperscript{rd} and 4\textsuperscript{th} rows) shows a strong response, however if the inverse \gls{PIAA} is omitted (3\textsuperscript{rd} row) the pattern is distorted and sensitivity is lost due to the higher sampling required.}\label{fig:diff_apert}
\end{figure}

First, we qualitatively evaluate the performance of the system for different apertures.\ \autoref{fig:diff_apert} illustrates the optimised system for the Magellan, \gls{GMT} and \gls{ELT} apertures. In the top row, we visualise the amplitude of the electric field after the first pair of \gls{PIAA} lenses (\autoref{fig:different_apertures_radial_profiles} shows an alternative visualisation with radial profiles). As expected, the apodisation concentrates the starlight towards the centre of the aperture. In the case of a clear aperture these would approximate prolate spheroids within manufacturing freedoms. For non-circularly symmetric apertures, however, the shape is clearly distorted. In the second row we depict the reference intensity (response of the system to a flat wavefront) $I_0$. All apertures share the behaviour where light from the geometric pupil is diffracted out into a circle that circumscribes it. There is a weak halo from the light diffracted by the mask, as is the case for the conventional \gls{ZWFS}. This is also the reason for the reduction of sensitivity to low frequencies as seen in \autoref{fig:max_diff_freqs}. In the third row we illustrate the response of the wavefront sensor to a high frequency Fourier mode at the pupil, both without and with a pair of inverse \gls{PIAA} lenses. Without the inverse optics, the pupil geometry remains distorted and contours of the intensity change are no longer vertical. Near the centre of the pupil these changes are more concentrated, requiring more pixels to sample the changes. This would reduce the sensitivity of the sensor in the faint limit, where read noise dominates per pixel. On the other hand, the system with an inverse \gls{PIAA} (last row) has restored geometry and uniform sampling. 

\subsubsection{Stellar magnitude}
\begin{figure}[h]
    \centering
    \includegraphics{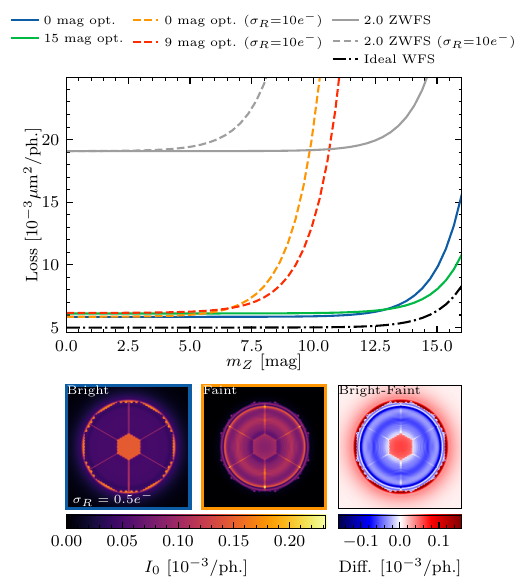}
    \caption{PIAA-ZWFS performance on the ELT (240 pixels across the pupil diameter).\ \textit{Top:} Loss (average variance of the maximum-likelihood estimate) normalised to photon count as a function of stellar magnitude. Different lines correspond to designs optimised for different read noise and/or stellar magnitude. The ideal WFS has the same read noise as the other sensors with solid lines ($0.5e^-$).\ \textit{\nobreak Bottom:} reference intensities for the bright (left) and faint (middle) design, along with the difference of the two (right). The faint design has put the flux into fewer pixels, particularly outside of the pupil, minimising the effect of read noise.}\label{fig:mag_change}
\end{figure}

Next, in \autoref{fig:mag_change}, we evaluate the \gls{PIAA}-\gls{ZWFS} with varying flux, and optimise the design for both bright and faint operation. We continue to use the loss metric (average variance of a maximum-likelihood estimator) normalised per photon to observe the underlying trend. We simulate both a photon counting camera (read noise 0.5$e^-$ \gls{RMS}) and a low quality camera (read noise 10$e^-$ \gls{RMS}) running at 1kHz with an end-to-end system throughput of 30\%. 

The shape of the loss function with decreasing flux is flat in the bright regime, before sharply increasing at an elbow. Optimising the system anywhere in the flat region (such as the 0 or 10 magnitude optimisation for $\sigma_R=0.5e^-$ \gls{RMS}) leads to the same result. This means that a single \gls{PIAA}-\gls{ZWFS} should suffice for all bright stars. After the turn-off, however, it is possible to optimise the system for significantly better performance, gaining about 0.75 magnitudes. The reference intensities (bottom) behave as expected: in the faint case, starlight is concentrated, with less diffraction outside the geometric pupil.

Even though \autoref{eq:fi} shows that changing overall flux is not the same as changing read noise, the behaviour for a higher read noise camera is qualitatively very similar, with the performance degradation occurring about $2.5\log_{10}((10/0.5)^2)\approx 6.5$ magnitudes earlier. 

We note that this analysis assumed a constant number of modes between bright and faint modes for a valid comparison. However, in order to fully optimise a system for faint operation the number of modes corrected and the number of pixels read should be carefully considered in the context of the requirements of the adaptive optics system.

\subsubsection{Bandwidth}\label{sec:bw}
An important factor of any wavefront sensor is how the performance changes with bandwidth, since a large bandwidth is ultimately what provides photons, enabling \gls{AO} systems to run at fast speeds and on fainter targets. In this experiment we keep the number of photons per frame fixed (and in the photon-noise-limited regime) and change the fractional bandwidth of the input field. 
\begin{figure}[h]
    \centering
    \includegraphics{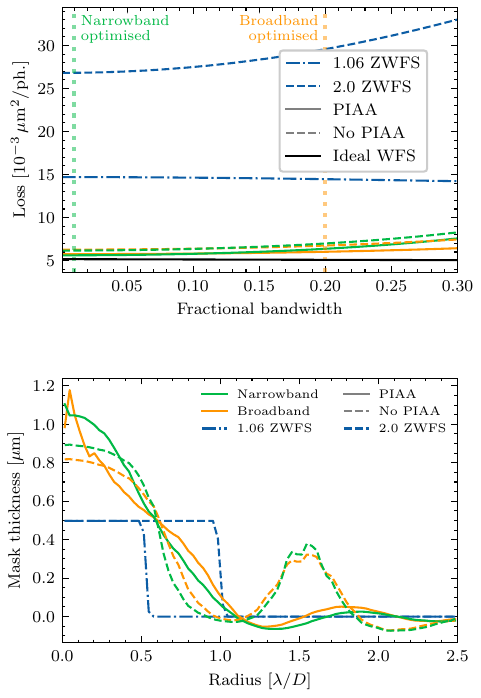}
    \caption{
    Wavefront sensor performance for different bandwidth on the Magellan aperture.\ \textit{Top}: The average variance of the prediction as a function of changing bandwidth. A Zernike \gls{WFS} with dot size 2$\lambda/D$ (2.0 ZWFS, blue dashed) shows a large (23\% at 0.3 bandwidth compared to monochromatic) degradation of performance with increasing bandwidth. We illustrate versions of our designs optimised for 20\% (orange) and 1\% (green) bandwidth. Designs optimised for larger bandwidths degrade less when compared to the 2.0 ZWFS. Designs without a \gls{PIAA} (dashed) degrade faster and perform worse. The PIAA-ZWFS designed for broadband operation only performs 12\% worse when compared to the monochromatic case. An ideal, achromatic wavefront sensor (black) performs well at all bandwidths. \textit{Bottom}: The phase masks from the optimised designs. The addition of the PIAA results in a phase mask with less structure beyond 1$\lambda/D$, explaining the better robustness of chromatic effects. Optimising for a larger bandwidth increases the size of the dot slightly and makes it more uniform, except at the very centre. }
    \label{fig:bandwidth_dep}
\end{figure}

\autoref{fig:bandwidth_dep} illustrates how the performance of the wavefront sensors studied degrades with wider bandwidth ($\Delta\lambda/\lambda_c$, where $\Delta\lambda$ is the width of the observing filter). We optimise four different designs: a combination of monochromatic vs. 0.2 fractional bandwidth, and no \gls{PIAA} vs. \gls{PIAA}. All optimised designs outperform conventional Zernike wavefront sensors and approach the ideal wavefront sensor bound in the narrowband limit.

Inspecting the converged phase mask designs in the bottom of \autoref{fig:bandwidth_dep}, we find that the inclusion of a \gls{PIAA} results in the removal of mask rings in the wings of the diffraction pattern, explaining the reduction of chromatic behaviour. Negative values represent a thickness that is below that of the initial substrate.

Ultimately, our results demonstrate that, whilst not completely insensitive to chromatic effects, the PIAA-ZWFS can be designed for robustness to different wavelengths, even when limited to fused silica alone. Future work can address more complex designs, such as doublet lenses or reflective optics, possibly with Wyne correctors \citep{wyne1979extending}.

\subsubsection{Stellar size}
With the coming generation of \glspl{ELT}, some scientifically interesting targets can no longer be assumed to be point sources. We explore the effect of stellar size on different wavefront sensors, and design PIAA-ZWFS designs optimised for resolved targets. We modify the simulator to inject a uniform disk as the source on sky by incoherently integrating over a range of tip/tilt values (approximated as a sum over 3 radial samples and 6 angular samples on a polar grid). This is done in the photon-limited regime as these targets are also bright, with the \gls{ELT} aperture for a wavefront sensor operating in the Z band.

\begin{figure}[t!]
    \centering
    \includegraphics[scale=0.97]{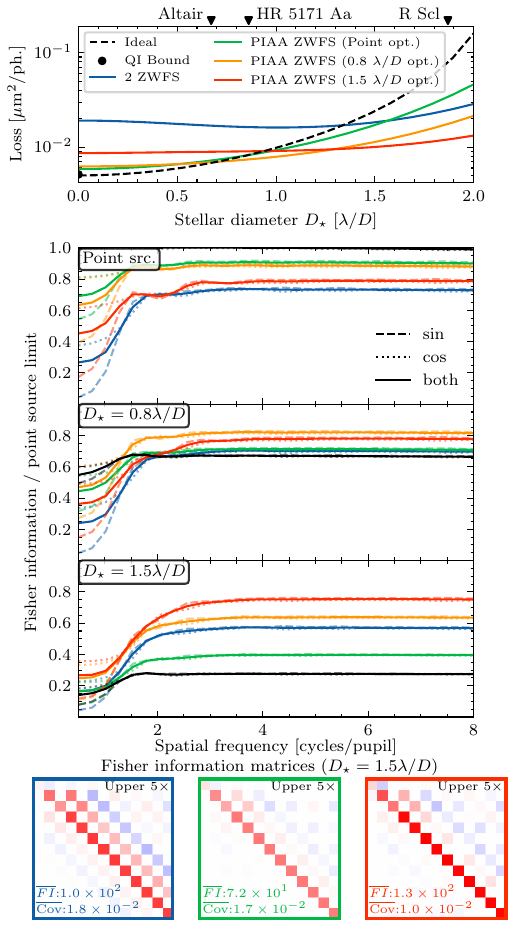}
    \caption{
    Resolved objects on the ELT, Z band. \textit{Top:}~\gls{WFS} performance as a function of stellar size. Performance declines with coherence, notably for the ideal \gls{WFS} (black, dashed) which meets the quantum information (QI) bound (black dot) for point sources. Our optimised designs for resolved objects (orange and red) outperform the ``ideal'' system for $D_\star=0.8\lambda/D$. 
    \textit{\nobreak Middle:}~Fisher Information (relative to the point source limit) vs. spatial frequency for each \gls{WFS} at a different stellar size;  solid lines illustrate the quadrature average of sine/cosine terms. Only the resolved-source-optimised PIAA-ZWFS resists sensitivity loss as stellar size increases.
    \textit{Bottom:}~Fisher information matrices for the 2 ZWFS (blue), point-optimised PIAA-ZWFS (green) and the resolved-source-optimised (1.5$\lambda/D$) PIAA-ZWFS (red), with a common colour scale. The upper half has values inflated by a factor 5 to show detail. Insets state the average Fisher information (FI) and mean variance (Cov) per photon. While the 2 ZWFS has larger values on the diagonal than the point-optimised PIAA-ZWFS, the cross correlation terms (non-zero off diagonal) result in the same variance value. The resolved-source-optimised design has both high information and low cross terms.
    }
    \label{fig:stellar_size_dep}
\end{figure}

\autoref{fig:stellar_size_dep} shows the results from this experiment. The top plot depicts the performance of the wavefront sensor on a logarithmic scale as a function of stellar diameter $D_\star$. Almost all wavefront sensors perform worse as the stellar size grows, with the exception of the 2 ZWFS, which begins with relatively poor performance. In particular, the ``ideal'' wavefront sensor -- which is quantum optimal for point sources -- rapidly becomes sub-optimal. Our approach is able to outperform it for $D_\star>0.8\lambda/D$. We hypothesise that this rapid degradation is because the ideal wavefront sensor is sensitive to tip/tilt modes, and exploits coherence which these larger sources do not have. The middle plots show the information behaviour as a function of spatial frequency. The information gain of the ideal wavefront sensor reduces approximately equally at all frequencies, while designs that are optimised for resolved sources (orange and red) maintain consistent behaviour for source sizes smaller than the design value. All wavefront sensors struggle to constrain tip/tilt like modes (dotted lines at low frequencies) when the source becomes resolved.

The middle plots are commonly used in the literature (usually plotting the square root of the information) to support one wavefront sensor over another. In the bottom plot, we show an example where this visualisation has the potential to obscure wavefront sensing performance where multiple aberrations are present. Looking at the $D_\star=1.5\lambda/D$ case alone, one might come to the conclusion that the 2 ZWFS (blue) is equal to or more sensitive than the point-optimised PIAA-ZWFS (green) at all frequencies. If we inspect the Fisher information matrices, we do indeed observe that the 2 ZWFS diagonal has almost 40\% larger values than the PIAA-ZWFS, however this also contains significant cross correlation off the diagonal. This means that the responses of the wavefront sensor to orthonormal modes are not orthogonal, and that errors have the capacity to couple between different aberration coefficients. This is reflected in the average of the diagonal elements of the (not pictured) covariance matrix ($\overline{\text{Cov}}$), for which the PIAA-ZWFS is marginally better. The resolved-source-optimised PIAA-ZWFS (red) maintains high information content along the diagonal whilst also minimising off-diagonal components. While the discrepancy between the sensitivity plots and full Fisher matrix is most clearly seen for the resolved source case, we note that this behaviour occurs in any system.

\subsection{Ablation study}
We now turn to a direct comparison between different sensors in a typical case: a Z band wavefront sensor on the Magellan telescope with 60 actuators across the pupil diameter in the photon limited regime. To highlight the importance of each component shown in \autoref{fig:overview}, we run optimisations with all possible combinations of system architecture in addition to the standard 1.06 ZWFS and proposed 2 ZWFS. 

\begin{table}[h]
\caption{Ablation study for our optimised system. } 
\label{tab:ablation}
\centering
\begin{tabular}{@{}ccccc@{}}
\toprule
System    & PIAA          & AM  & Loss ($10^{-3}\, \downarrow$)     & Sub-opt. (\%) \\ \midrule
Ideal  &               &               & \textbf{5.12} &                  \\ 
1.06 ZWFS &               &               & 14.5 & 183             \\
2.0 ZWFS  &               &               & 29.5 & 475             \\ \hline
Ours      &               &               &     7.35 & 44                 \\
Ours      & \boldcheckmark &               &     6.29 & 23               \\
Ours      &               & \boldcheckmark & 6.72 & 31            \\
Ours      & \boldcheckmark & \boldcheckmark & \textbf{6.01} & 17           \\ \bottomrule
\end{tabular}
\tablefoot{AM: annular mask. Loss (normalised average variance of a maximum-likelihood estimator) and sub-optimality as a fraction of an ideal wavefront sensor $100(L/L_{\mathrm{ideal}}-1)\%$. The use of \gls{PIAA} lenses and an annular mask instead of a simple dot each provide improvements.}
\end{table}

\autoref{tab:ablation} contains the results of the ablation study for our system. In the experiments of our design without the annular mask, the phase dot can be varied in diameter and thickness. Results are presented as absolute loss (the average variance per mode per photon of a maximum-likelihood estimator) and the sub-optimality of the sensor as a percentage of the loss of an ideal wavefront sensor. We note that reflections are not accounted for as this can depend on manufacturing decisions, but due to the normalisation of the loss by the flux the effect can be easily accounted for. For example, assuming a Thorlabs `B' coating with an average reflectance of 0.26\% per surface, the resulting loss for the PIAA and annular mask system (14 surfaces) is $6.23\times10^{-3}$ while the annular mask only (6 surfaces) system has a loss of $6.83\times10^{-3}$. Hence the use of the PIAA lenses remains a net positive. 

From these results we identify that the PIAA-ZWFS closes the gap between the ideal wavefront sensor and the 1.06 ZWFS by around an order of magnitude (17\% sub-optimality rather than 183\%), and by a factor 2.5 compared to an optimised \gls{ZWFS}. We do, however, note that the exact value depends on the basis selected and the number of high frequency modes.

\subsection{Wavefront reconstruction}
We now validate our optimisation approach and explore the dynamic range of the \gls{PIAA}-\gls{ZWFS} by reconstructing aberrated wavefronts. We generate different turbulence \gls{OPD} screens with a power law (exponent $-5/3$, expected for a second stage after a servo-lag dominated system), project them onto a basis and then try to reconstruct the basis terms. We also present results for a power law exponent of $-11/3$ in \autoref{fig:recon_errors_binned_113}. Errors are only measured relative to the aberrations that are spanned by the basis set. To show the flexibility of the above optimisation, we use the same optical models as from the ablation study, changing only the \gls{OPD} basis to now consider the first 45 Zernike modes (Noll indexed, excluding piston) rather than a one dimensional Fourier basis. We show the reconstructor performance in a photon limited regime and add Poisson noise to the data.

Linear reconstructors are created by using the same auto-differentiation framework to calculate the interaction matrix. The nonlinear reconstructors are differentiable forward models. Initialised from a flat phase at the pupil, these models fit modal coefficients using \gls{BFGS}, minimising the negative logarithm of the probability mass function for the Poisson distribution. While this is impractical for real time control, it shows the realistic dynamic range, identifying the aberration region where real-time nonlinear reconstructors (such as neural networks) should focus.

\begin{figure}[h]
    \centering
    \includegraphics{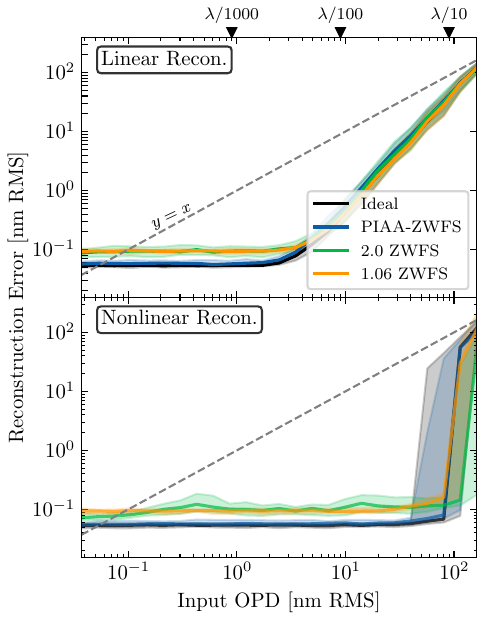}
    \caption{
    Wavefront estimate errors under varying input aberration size for different wavefront sensors and reconstructors for a Zernike aberration basis drawn from a power law with exponent $-5/3$. Results are binned by input \gls{OPD}, with the line showing the median and the shaded area showing the 16-84 percentiles of each bin. The black dashed line indicates no reconstruction ($y=x$). The PIAA-ZWFS outperforms the classical \gls{ZWFS} ($1.06\lambda/D$ dot diameter) and the $2.0$ variant. It is comparable with the ideal point source wavefront sensor. The performance of all linear reconstructors degrades equally fast with nonlinear behaviours. A nonlinear reconstructor performs well for all sensors and extends the range by over an order of magnitude.
    }
    \label{fig:recon_errors}
\end{figure}

\autoref{fig:recon_errors} shows the reconstructor error over the aperture for different sensors using a linear (top) and nonlinear (bottom) reconstruction method in the Z band ($\lambda_c=900$\,nm). The results are binned and the median is presented as the line with the 16-84 percentile shaded. In the low aberration case, the PIAA-ZWFS is comparable with an ideal wavefront sensor while both the classical 1.06 \gls{ZWFS} and $2.0$ variant perform worse. This is as expected from the optimisation results, although a one-to-one correspondence in reconstruction error (from e.g. \autoref{tab:ablation}) is not expected due to changes in the sensing basis used. As the aberrations grow larger, the nonlinear behaviour appears in all sensors before the 10\,nm ($\lambda/90$) RMS point, growing until there is no correct estimate after 100\,nm ($\lambda/9$) RMS. Using a nonlinear reconstructor, however, we can extend the range of photon noise limited estimates by an order of magnitude, before catastrophic failures of the reconstruction emerge around 100\,nm ($\lambda/9$) RMS. This point appears marginally earlier for the ideal wavefront sensor and the PIAA-ZWFS compared to the classical \gls{ZWFS}. These large aberrations result in phase wrapping, where the aberration producing the sensor image is not the same local minimum in common with the initialisation point. In the large aberration regime with many modes choosing the right initialisation point becomes intractably difficult. As this simulation is conducted with a central wavelength of 900\,nm, we expect peak-to-valley changes of order $2\pi$ radians at around 100\,nm \gls{RMS}. In our experience, varying the basis set tweaks the exact failure point (whilst maintaining the same overall shape in both linear and nonlinear reconstruction), indicating that dynamic range weakly depends on the aberration distribution.

\section{Discussion}

It is evident that the PIAA-ZWFS effectively closes the gap towards the fundamental sensing limit in the high-Strehl regime. The degrees of freedom given in the design space (namely \gls{PIAA} sag and annular rings of the phase mask) enable a family of designs that can be optimised for different apertures, flux levels, aberration modes, bandwidths and source distributions. The increased sensitivity to phase aberrations compared to the classical \gls{ZWFS} is due to mode matching, where the reference beam is shaped to match that of a beam from a pinhole the size of the phase mask. 
The use of PIAA lenses makes the phase mask less extended, decreasing the bandwidth sensitivity of the designs and easing fabrication area requirements. 

In early experiments we were able to replicate a saturation of the sensitivity bound for high spatial frequencies when optimising for the average Fisher information and when sampling the detector with enough pixels, first observed in \cite{haffert2023reaching}. However, in all adaptive optics systems, phase aberrations do not occur at only a single frequency. It is not the sensitivity/information gain of any particular mode that matters, but rather the uncertainty on the reconstruction solution of a maximum-likelihood estimator that determines the performance (in, e.g. Strehl ratio). Hence, our optimisation metric of the trace of the covariance matrix correctly manages correlations between responses to different orthogonal phase aberrations, and weighs the sensitivity appropriately when optimising for a generic quantity such as Strehl ratio. This behaviour is most clearly seen in the last row of \autoref{fig:stellar_size_dep}, where our resolved-source-optimised design has both large information content and low average variance (via low crosstalk), particularly compared to the 2\gls{ZWFS}. The validity of our metric is also reinforced by the reconstruction results shown in \autoref{fig:recon_errors}, where the optimised design approaches an ideal wavefront sensor and outperforms the \gls{ZWFS} and 2\gls{ZWFS} in the photon and low aberration limit. 

For true end-to-end optimisation, the covariance matrix can be fed into an application specific loss function--we do not do this, instead showing the system as a whole for a generic task. Priors can be easily incorporated by adding a matrix to the Fisher information matrix before inversion. Another generic metric could be the log determinant of the covariance matrix (corresponding to minimising the volume of the confidence ellipsoid); we found very similar results for either metric on preliminary experiments and decided the average variance was easier to interpret. 

Whilst significantly closer to the information limit than other designs explored here, the PIAA-ZWFS is still sub-optimal, most noticeably for low order modes (\autoref{fig:stellar_size_dep}, point source case), although it still outperforms existing methods. This sub-optimality is related to the light scattered outside the pupil observed in \autoref{fig:diff_apert}. As the beam is tilted off the phase mask, the amount of light scattered is reduced. This decrease, however, does not contribute to the information gained as the sign of the change is ambiguous. For higher spatial frequencies, this is effectively a change in the intensity of the reference beam, which only changes quadratically with the applied aberration. In existing systems, however, the light outside the pupil is still practically useful, providing telemetry information and an estimate of the on-axis Strehl ratio. A truly ideal wavefront sensor would not have access to this additional debugging information. 

Recent work by \cite{trzaska2026fundamental} presented the \gls{PAWS}, a sensor design that can, in principle, phase shift the piston mode alone by $\pi/2$ (and hence is quantum optimal in the high-Strehl regime). The system presented requires at least 42 glass surfaces and 4 mirrors, compared to just 14 glass surfaces for the PIAA-ZWFS presented here. Assuming the same coating as before for the glass surfaces and a standard 2\% loss for the mirrors, the PAWS loss (in the context of \autoref{tab:ablation} as the ideal wavefront sensor) becomes $6.19\times10^{-3}$ (from $5.12\times10^{-3}$), compared to the reflection loss adjusted value of $6.23\times10^{-3}$ calculated before for the PIAA-ZWFS. These are comparable at the level of implementation details and the choice of basis as discussed previously. Furthermore, we believe the PIAA-ZWFS is significantly easier to manufacture (as it does not require broadband, achromatic phase optics) and align (as there is only one focal plane optic instead of four) than the \gls{PAWS}, however implementation in hardware is left as future work. 

Finally, we consider the importance of the dynamic range of the wavefront sensor presented. Despite the gain in sensitivity provided by the PIAA-ZWFS, dynamic range is not lost in linear reconstructors relative to the classical \gls{ZWFS}, nor nonlinear reconstructors relative to an ideal wavefront sensor. Hence, we believe that the same reconstruction techniques that are being developed for the \gls{ZWFS} apply to the PIAA-ZWFS. As a second-stage wavefront sensor, the sensing requirements on reconstruction depend on the first stage residuals and the parameters of the second-stage control loop. If operating in the nonlinear region, different strategies are possible. We note that a closed loop system could converge even with the linear reconstructor, given the timescale of the evolution is slower than the convergence time. Another option is to leverage fast nonlinear reconstructors, such as neural networks which have been demonstrated on-sky for pyramid wavefront sensors \citep{landman2025making}. Finally, there is also the possibility to alter the hardware, capturing sensor frames in multiple wavelengths and/or polarisations simultaneously. Multiple wavelengths have been shown to improve dynamic range significantly in simulation \citep{darcis2025adding} and also have good synergies with emerging technologies such as microwave kinetic inductance detectors \citep{mazin2019optical}. Similarly, a vector \gls{ZWFS} has been shown in a lab setting to admit a much larger dynamic range \citep{doelman2018simultaneous, wenger2025polarization}. The optimisation framework presented in this work can be easily adapted to these cases.

\section{Conclusions}

Second-stage wavefront sensors are a critical part of extreme \gls{AO} systems in high-contrast and high angular resolution imaging. Sensors that provide the most information about phase aberrations per photon enable such systems to provide better correction by running at faster speeds or on fainter targets. Similarly, these sensors would significantly reduce the time needed to reach picometre knowledge for high-contrast instruments in space. While the information limits of wavefront sensing have been known for some time, the search for architectures that reach these limits is ongoing. The PIAA-ZWFS presented in this work makes significant strides towards this target. 

While it is possible to optimise the PIAA-ZWFS architecture to saturate the limit at high spatial frequencies, we instead stress the importance of using the variance of an unbiased, maximum-likelihood estimator as the metric for performance. This choice more accurately reflects the closed loop residuals and correctly accounts for correlations between phase aberrations. Through this framework we optimise the PIAA-ZWFS in a differentiable simulator. Our results demonstrate the sensor functions well on a range of apertures, bandwidths, fluxes and stellar sizes. Our sensor outperforms the ``ideal'' wavefront sensor on extended objects with diameter larger than $0.8\lambda/D$. Our ablation study demonstrates that each complexity added to the system improves performance significantly. For the basis selected, we close the gap to the ideal wavefront sensor by an order of magnitude compared to the previous state-of-the-art sensor -- the \gls{ZWFS}. Finally, we verify that the sensor's dynamic range is not drastically reduced by the gain in sensitivity through reconstruction of a mixture of phase aberrations. 

There are several exciting avenues of future work. Most critically, hardware development and verification of the simulation results presented in this work are needed. Different processes can be employed to manufacture the phase masks presented in this work, such as greyscale lithography. A less explored process would involve a metasurface phase mask, which could, in principle, enable wavefront sensing and coronagraphy with a single optic \citep{patel2026demonstration}. As part of the hardware development an understanding of the practicalities of the system is important, namely how the system degrades with misalignment and manufacturing imperfections. Another practical avenue of future work is to explore how, given a list of desired targets, different operating modes of a manufactured PIAA-ZWFS could be designed while holding the \gls{PIAA} lenses common among all modes. Finally, the dynamic range of the sensor should be tested empirically, with both software (such as nonlinear, neural network reconstructors) and hardware (such as multiple wavelength images) modifications.

\begin{acknowledgements}
The authors acknowledge funding from Dutch Research Council (NWO) VI.Vidi.233.144. This work was performed using the compute resources from the Academic Leiden Interdisciplinary Cluster Environment (ALICE) provided by Leiden University. We would like to thank R. Landman, M. Ireland, M. Mars and J. K. Wallace for helpful discussions.

Microsoft Copilot was used in developing the software for this work. OpenAI Codex was used to modify code for figure generation in the revision stage.
This work was supported in part by a grant of access to OpenAI models through the ChatGPT for Academic Researchers program.
\end{acknowledgements}

\bibliographystyle{bibtex/aa.bst} 
\bibliography{refs}

@article{paterson2008towards,
  title={Towards practical wavefront sensing at the fundamental information limit},
  author={Paterson, C},
  journal={Journal of Physics: Conference Series},
  volume={139},
  number={1},
  pages={012021},
  year={2008},
  organization={IOP Publishing}
}

@article{hoffnagle2000design,
  title={Design and performance of a refractive optical system that converts a Gaussian to a flattop beam},
  author={Hoffnagle, John A and Jefferson, C Michael},
  journal={Applied Optics},
  volume={39},
  number={30},
  pages={5488--5499},
  year={2000},
  publisher={Optical Society of America}
}

@article{guyon2018extreme,
  title={Extreme adaptive optics},
  author={Guyon, Olivier},
  journal={ARA\&A},
  volume={56},
  number={1},
  pages={315--355},
  year={2018},
  publisher={Annual Reviews}
}

@article{chambouleyron2023modeling,
  title={Modeling noise propagation in Fourier-filtering wavefront sensing, fundamental limits, and quantitative comparison},
  author={Chambouleyron, Vincent and Fauvarque, Olivier and Plantet, C{\'e}dric and Sauvage, J-F and Levraud, Nicolas and Ciss{\'e}, Mahawa and Neichel, Beno{\^\i}t and Fusco, Thierry},
  journal={A\&A},
  volume={670},
  pages={A153},
  year={2023},
  publisher={EDP Sciences}
}

@article{wyne1979extending,
  title={Extending the bandwidth of speckle interferometry},
  author={Wyne, CG},
  journal={Optics Communications},
  volume={28},
  number={1},
  pages={21--25},
  year={1979},
  publisher={Elsevier}
}

@article{bracewell1966fourier,
  title={The Fourier transform and its applications},
  author={Bracewell, Ron and Kahn, Peter B},
  journal={American Journal of Physics},
  volume={34},
  number={8},
  pages={712--712},
  year={1966},
  publisher={American Association of Physics Teachers}
}

@article{ragazzoni1999sensitivity,
  title={Sensitivity of a pyramidic wave front sensor in closed loop adaptive optics},
  author={Ragazzoni, Roberto and Farinato, J},
  journal={A\&A},
  volume={350},
  pages={L23--L26},
  year={1999}
}

@article{desdoigts2023differentiable,
  title={Differentiable optics with $\partial$Lux: {I}—deep calibration of flat field and phase retrieval with automatic differentiation},
  author={Desdoigts, Louis and Pope, Benjamin JS and Dennis, Jordan and Tuthill, Peter G},
  journal={Journal of Astronomical Telescopes, Instruments, and Systems},
  volume={9},
  number={2},
  pages={028007--028007},
  year={2023},
  publisher={Society of Photo-Optical Instrumentation Engineers}
}

@article{Desdoigts2026Differentiable,
	author = {Desdoigts, Louis and Pope, Benjamin and Gully-Santiago, Michael and Tuthill, Peter},
	journal = {The Open Journal of Astrophysics},
	doi = {10.33232/001c.172540},
	year = {2026},
	month = {oct 15},
	publisher = {Maynooth Academic Publishing},
	title = {Differentiable {Optics} with {dLux} {II}: Optical {Design} {Maximising} {Fisher} {Information}},
	volume = {9},
}

@article{trzaska2026fundamental,
  title={Fundamental limits to phase and amplitude estimation in the high-Strehl regime},
  author={Trzaska, Jacob and Ashok, Amit},
  journal={Journal of the Optical Society of America A},
  volume={43},
  number={6},
  pages={891--900},
  year={2026},
  publisher={Optica Publishing Group}
}

@article{galicher2005laboratory,
  title={Laboratory Demonstration and Numerical Simulations of the Phase-Induced Amplitude Apodization},
  author={Galicher, Raphael and Guyon, Olivier and Otsubo, Masashi and Suto, Hiroshi and Ridgway, Stephen},
  journal={PASP},
  volume={117},
  number={830},
  pages={411--420},
  year={2005},
  publisher={The University of Chicago Press}
}

@inproceedings{por2018high,
  title={High Contrast Imaging for Python (HCIPy): an open-source adaptive optics and coronagraph simulator},
  author={Por, Emiel H and Haffert, Sebastiaan Y and Radhakrishnan, Vikram M and Doelman, David S and Van Kooten, Maaike and Bos, Steven P},
  booktitle={Adaptive Optics Systems VI},
  volume={10703},
  pages={1112--1125},
  year={2018},
  organization={SPIE}
}

@article{bradbury2021jax,
  title={Jax: Autograd and {XLA}},
  author={Bradbury, James and Frostig, Roy and Hawkins, Peter and Johnson, Matthew James and Leary, Chris and Maclaurin, Dougal and Necula, George and Paszke, Adam and VanderPlas, Jake and Wanderman-Milne, Skye and others},
  journal={Astrophysics Source Code Library},
  pages={ascl--2111},
  year={2021}
}

@article{slepian1965analytic,
  title={Analytic solution of two apodization problems},
  author={Slepian, David},
  journal={Journal of the Optical Society of America},
  volume={55},
  number={9},
  pages={1110--1115},
  year={1965},
  publisher={Optical Society of America}
}

@article{guyon2003phase,
  title={Phase-induced amplitude apodization of telescope pupils for extrasolar terrestrial planet imaging},
  author={Guyon, Olivier},
  journal={A\&A},
  volume={404},
  number={1},
  pages={379--387},
  year={2003},
  publisher={EDP Sciences}
}

@inproceedings{chambouleyron2022optimizing,
  title={Optimizing Fourier-filtering WFS to reach sensitivity close to the fundamental limit},
  author={Chambouleyron, Vincent and Fauvarque, Olivier and Plantet, C{\'e}dric and Sauvage, J-F and Levraud, Nicolas and Ciss{\'e}, Mahawa and Neichel, Beno{\^\i}t and Fusco, Thierry},
  booktitle={Adaptive Optics Systems VIII},
  volume={12185},
  pages={932--944},
  year={2022},
  organization={SPIE}
}

@article{kasper2021pcs,
  title   = {{PCS}---A Roadmap for Exoearth Imaging with the {ELT}},
  author  = {Kasper, Markus and Cerpa Urra, Nelly and Pathak, Prashant and Bonse, Markus and Nousiainen, Jalo and Engler, Byron and Heritier, C{\'e}dric Ta{\"i}ssir and Kammerer, Jens and Leveratto, Serban and Rajani, Chang and Bristow, Paul and Le Louarn, Miska and Madec, Pierre-Yves and Str{\"o}bele, Stefan and Verinaud, Christophe and Glauser, Adrian and Quanz, Sascha P. and Helin, Tapio and Keller, Christoph and Snik, Frans and Boccaletti, Anthony and Chauvin, Ga{\"e}l and Mouillet, David and Kulcs{\'a}r, Caroline and Raynaud, Henri-Fran{\c{c}}ois},
  journal = {The Messenger},
  number  = {182},
  pages   = {38--43},
  year    = {2021},
  month   = mar,
  doi     = {10.18727/0722-6691/5221}
}

@article{n2013calibration,
  title={Calibration of quasi-static aberrations in exoplanet direct-imaging instruments with a Zernike phase-mask sensor},
  author={N’Diaye, Mamadou and Dohlen, K and Fusco, T and Paul, B},
  journal={A\&A},
  volume={555},
  pages={A94},
  year={2013},
  publisher={EDP Sciences}
}

@article{guyon2005limits,
  title={Limits of adaptive optics for high-contrast imaging},
  author={Guyon, Olivier},
  journal={ApJ},
  volume={629},
  number={1},
  pages={592},
  year={2005},
  publisher={IOP Publishing}
}

@inproceedings{patel2026demonstration,
  title={Demonstration of simultaneous PIAA-coronagraphy and wavefront sensing using a single metasurface-based focal-plane optic},
  author={Patel, Dhwanil and Haffert, Sebastiaan Y and Palatnick, Skyler and Taras, Adam and Millar-Blanchaer, Maxwell and Mars, Matthjis and Tonucci, Elena and Males, Jared R and Close, Laird M and Liberman, Joshua and others},
  booktitle={Advances in Optical and Mechanical Technologies for Telescopes and Instrumentation VII},
  volume={14154},
  pages={1367--1377},
  year={2026},
  organization={SPIE}
}

@article{lindley1956measure,
  title={On a measure of the information provided by an experiment},
  author={Lindley, Dennis V},
  journal={The Annals of Mathematical Statistics},
  volume={27},
  number={4},
  pages={986--1005},
  year={1956},
  publisher={Institute of Mathematical Statistics}
}

@article{rainforth2024modern,
  title={Modern Bayesian experimental design},
  author={Rainforth, Tom and Foster, Adam and Ivanova, Desi R and Bickford Smith, Freddie},
  journal={Statistical Science},
  volume={39},
  number={1},
  pages={100--114},
  year={2024},
  publisher={Institute of Mathematical Statistics}
}

@article{jovanovic2017efficient,
  title={Efficient injection from large telescopes into single-mode fibres: Enabling the era of ultra-precision astronomy},
  author={Jovanovic, N and Schwab, C and Guyon, O and Lozi, J and Cvetojevic, N and Martinache, Frantz and Leon-Saval, S and Norris, B and Gross, S and Doughty, D and others},
  journal={A\&A},
  volume={604},
  pages={A122},
  year={2017},
  publisher={EDP Sciences}
}

@mastersthesis{foster2023implementation,
  title={Implementation of Phase Induced Apodization Optics on MagAO-X},
  author={Foster, Warren Byrum},
  year={2023},
  school={The University of Arizona}
}

@article{tonucci2026phase,
  title={Phase-Induced Amplitude Apodization Complex Mask Coronagraph (PIAACMC) on-sky demonstration with MagAO-X},
  author={Tonucci, Elena and Haffert, SY and Foster, WB and Males, JR and Guyon, Olivier and Close, Laird M and Van Gorkom, Kyle and Hedglen, AD and Johnson, PT and Kautz, MY and others},
  journal={A\&A},
  volume={708},
  pages={A144},
  year={2026},
  publisher={EDP Sciences}
}

@article{steeves2020picometer,
  title={Picometer wavefront sensing using the phase-contrast technique},
  author={Steeves, John and Wallace, J Kent and Kettenbeil, Christian and Jewell, Jeffrey},
  journal={Optica},
  volume={7},
  number={10},
  pages={1267--1274},
  year={2020},
  publisher={Optical Society of America}
}

@article{wenger2025polarization,
  title={Polarization-dependent metasurface for vector Zernike wavefront sensing with increased dynamic range},
  author={Wenger, Tobias and Wallace, J Kent},
  journal={Optics Letters},
  volume={50},
  number={3},
  pages={726--729},
  year={2025},
  publisher={Optica Publishing Group}
}

@inproceedings{doelman2017patterned,
  title={Patterned liquid-crystal optics for broadband coronagraphy and wavefront sensing},
  author={Doelman, David S and Snik, Frans and Warriner, Nathaniel Z and Escuti, Michael J},
  booktitle={Techniques and Instrumentation for Detection of Exoplanets VIII},
  volume={10400},
  pages={224--235},
  year={2017},
  organization={SPIE}
}

@article{doelman2018simultaneous,
  title={Simultaneous phase and amplitude aberration sensing with a liquid-crystal vector-Zernike phase mask},
  author={Doelman, David S and Fagginger Auer, Fedde and Escuti, Michael J and Snik, Frans},
  journal={Optics Letters},
  volume={44},
  number={1},
  pages={17--20},
  year={2018},
  publisher={Optical Society of America}
}

@article{chang2018optical,
  title={Optical metasurfaces: progress and applications},
  author={Chang, Shengyuan and Guo, Xuexue and Ni, Xingjie},
  journal={Annual Review of Materials Research},
  volume={48},
  number={1},
  pages={279--302},
  year={2018},
  publisher={Annual Reviews}
}

@article{grushina2019direct,
  title={Direct-write grayscale lithography},
  author={Grushina, Anya},
  journal={Advanced Optical Technologies},
  volume={8},
  number={3-4},
  pages={163--169},
  year={2019},
  publisher={De Gruyter}
}

@article{landman2025making,
  title={Making the unmodulated pyramid wavefront sensor smart-II. First on-sky demonstration of extreme adaptive optics with deep learning},
  author={Landman, Rico and Haffert, Sebastiaan Y and Long, JD and Males, JR and Close, LM and Foster, WB and Van Gorkom, K and Guyon, O and Hedglen, Alexander D and Johnson, PT and others},
  journal={A\&A},
  volume={696},
  pages={L1},
  year={2025},
  publisher={EDP Sciences}
}

@article{mazin2019optical,
  title   = {Optical and Near-{IR} Microwave Kinetic Inductance Detectors ({MKIDs}) in the 2020s},
  author  = {Mazin, Benjamin A. and Bailey, Jeb and Bartlett, Jo and Bockstiegel, Clint and Bumble, Bruce and Coiffard, Gregoire and Currie, Thayne and Daal, Miguel and Davis, Kristina and Dodkins, Rupert and Fruitwala, Neelay and Jovanovic, Nemanja and Lipartito, Isabel and Lozi, Julien and Males, Jared and Mawet, Dimitri and Meeker, Seth and O'Brien, Kieran and Rich, Michael and Smith, Jenny and Steiger, Sarah and Swimmer, Noah and Walter, Alex and Zobrist, Nick and Zmuidzinas, Jonas},
  journal = {BAAS},
  volume  = {51},
  number  = {7},
  eid     = {17},
  year    = {2019}
}

@inproceedings{haffert2023reaching,
  title     = {Reaching the fundamental sensitivity limit of wavefront sensing on arbitrary apertures with the phase induced amplitude apodized Zernike wavefront sensor ({PIAA-ZWFS})},
  author    = {Haffert, Sebastiaan Y. and Males, Jared R. and Guyon, Olivier},
  booktitle = {Proceedings of the 7th Adaptive Optics for Extremely Large Telescopes Conference (AO4ELT7)},
  year      = {2023},
  address   = {Avignon, France},
  doi       = {10.13009/AO4ELT7-2023-081}
}

@article{zernike1934diffraction,
  title={Diffraction theory of the knife-edge test and its improved form, the phase-contrast method},
  author={Zernike, Frits},
  journal={MNRAS},
  volume={94},
  pages={377--384},
  year={1934}
}

@article{fauvarque2019kernel,
  title={Kernel formalism applied to Fourier-based wave-front sensing in presence of residual phases},
  author={Fauvarque, Olivier and Janin-Potiron, Pierre and Correia, Carlos and Br{\^u}l{\'e}, Yoann and Neichel, Benoit and Chambouleyron, Vincent and Sauvage, Jean-Francois and Fusco, Thierry},
  journal={Journal of the Optical Society of America A},
  volume={36},
  number={7},
  pages={1241--1251},
  year={2019},
  publisher={Optical Society of America}
}

@article{darcis2025adding,
  title={Adding colour to the Zernike wavefront sensor: Advantages of including multi-wavelength measurements for wavefront reconstruction},
  author={Darcis, M and Haffert, SY and Chambouleyron, V and Doelman, DS and de Visser, PJ and Kenworthy, MA},
  journal={A\&A},
  volume={701},
  pages={A157},
  year={2025},
  publisher={EDP Sciences}
}

@article{soummer2007fast,
  title={Fast computation of Lyot-style coronagraph propagation},
  author={Soummer, Remi and Pueyo, Laurent and Sivaramakrishnan, Anand and Vanderbei, Robert J},
  journal={Optics Express},
  volume={15},
  number={24},
  pages={15935--15951},
  year={2007},
  publisher={Optical Society of America}
}

@article{chambouleyron2024coronagraph,
  title={Coronagraph-based wavefront sensors for the high Strehl regime},
  author={Chambouleyron, Vincent and Wallace, J Kent and Jensen-Clem, Rebecca and Macintosh, Bruce},
  journal={Optics Express},
  volume={32},
  number={27},
  pages={47706--47720},
  year={2024},
  publisher={Optica Publishing Group}
}

@article{forbes2007shape,
  title={Shape specification for axially symmetric optical surfaces},
  author={Forbes, Greg W},
  journal={Optics Express},
  volume={15},
  number={8},
  pages={5218--5226},
  year={2007},
  publisher={OSA}
}

@article{rader2024optimistix,
  title={Optimistix: modular optimisation in {JAX} and Equinox},
  author={Rader, Jason and Lyons, Terry and Kidger, Patrick},
  journal={arXiv preprint arXiv:2402.09983},
  year={2024}
}

@inproceedings{landman2022joint,
  title={Joint optimization of wavefront sensing and reconstruction with automatic differentiation},
  author={Landman, Rico and Keller, Christoph and Por, Emiel H and Haffert, Sebastiaan and Doelman, David and Stockmans, Thijs},
  booktitle={Adaptive Optics Systems VIII},
  volume={12185},
  pages={2582--2593},
  year={2022},
  organization={SPIE}
}

@book{fletcher2013practical,
  title={Practical methods of optimization},
  author={Fletcher, Roger},
  year={2013},
  publisher={John Wiley \& Sons}
}

@inproceedings{males_high-contrast_2024,
    address = {Yokohama, Japan},
    title = {High-contrast imaging at first-light of the {GMT}: the preliminary design of {GMagAO}-{X}},
    isbn = {978-1-5106-7515-5 978-1-5106-7516-2},
    shorttitle = {High-contrast imaging at first-light of the {GMT}},
    url = {https://www.spiedigitallibrary.org/conference-proceedings-of-spie/13096/3018157/High-contrast-imaging-at-first-light-of-the-GMT/10.1117/12.3018157.full},
    doi = {10.1117/12.3018157},
    language = {en},
    urldate = {2026-05-18},
    booktitle = {Ground-based and {Airborne} {Instrumentation} for {Astronomy} {X}},
    publisher = {SPIE},
    author = {Males, Jared R. and Close, Laird M. and Haffert, Sebastiaan Y. and Kautz, Maggie and Kelly, Douglas and Fletcher, Adam and Salanski, Thomas and Durney, Oliver and Noenickx, Jamison and Ford, John and Gasho, Victor and Pearce, Logan and Kueny, Jay and Guyon, Olivier and Weinberger, Alycia and Bowler, Brendan and Kraus, Adam and Batalha, Natasha},
    editor = {Vernet, Joël R. and Bryant, Julia J. and Motohara, Kentaro},
    month = jul,
    year = {2024},
    pages = {34},
}

@article{males_mysterious_2021,
    title = {The {Mysterious} {Lives} of {Speckles}. {I}. {Residual} {Atmospheric} {Speckle} {Lifetimes} in {Ground}-based {Coronagraphs}},
    volume = {133},
    issn = {0004-6280, 1538-3873},
    url = {https://iopscience.iop.org/article/10.1088/1538-3873/ac0f0c},
    doi = {10.1088/1538-3873/ac0f0c},
    language = {en},
    number = {1028},
    urldate = {2026-03-03},
    journal = {PASP},
    author = {Males, Jared R. and Fitzgerald, Michael P. and Belikov, Ruslan and Guyon, Olivier},
    month = oct,
    year = {2021},
    pages = {104504},
}

\begin{appendix}
    \nolinenumbers

\section{Derivation of Fisher information bounds}\label{app:derivs}

A general result for the derivative of the intensity with respect to a parameter is 
\begin{align}\label{eqn:chain}
    \frac{\partial_\beta I}{N_{\mathrm{ph}}} &= \partial_\beta \left(\left|E_o\right|^2\right) \\
    &= \partial_\beta \left(E_o^* E_o\right)\\
    &= \partial_\beta \left(E_o^*\right) E_o + E_o^*\partial_\beta \left(E_o\right)\\
    &= \partial_\beta \left(E_o\right)^* E_o + E_o^*\partial_\beta \left(E_o\right)\\
    &= 2 \Re\left\{E_o^* \partial_\beta E_o\right\}, \label{eq:diffI}
\end{align}
where we have used the fact that the derivative and complex conjugate are interchangeable.

\subsection{Sensitivity to photon noise}
This bound was previously demonstrated in \cite{paterson2008towards}. Here we present an alternative proof in a Fisher information framework.

This is a proof from first principles, starting from considering a single pixel with index $j$. We start by considering the measured intensity $\hat{I}_j$ (which is a random variable), such that the probability distribution of the pixel is given by a Poisson distribution
\begin{align}
    f_j(x_j | \theta) &= \text{Pr}(\hat{I}_j=x_j) = \frac{{I}_j^{x_j} \exp(-{I}_j)}{{x}_j!}.
\end{align}
The Fisher information with respect to a parameter $\theta$ then is
\begin{align}
    \text{FI}_{j} [\theta] &= \sum_{x_j=0}^\infty \frac{\left(\partial_\theta f_j(x_j | \theta)\right)^2}{f_j(x_j | \theta)}.
\end{align}
Applying the derivative and simplifying produces
\begin{align}
    \text{FI}_{j} [\theta] &= \left(\partial_\theta {I}_j \right)^2 \frac{1}{I_j^2} \sum_{x_j=0}^\infty \frac{{I}_j^{x_j} \exp(-{I}_j) (x_j-{I}_j)^2}{x_j!}\\
    &= \frac{\left(\partial_\theta {I}_j \right)^2}{{I}_j^2} \text{Var}[\hat{I}_j]\\
    &= \frac{\left(\partial_\theta {I}_j \right)^2}{{I}_j}.
\end{align}
This agrees with the expression previously derived by \cite{haffert2023reaching}. Note that for cross terms one can change the numerator to $\left(\partial_{\theta_1} {I}_j \right)\left(\partial_{\theta_2} {I}_j \right)$.

\textbf{Proof of the bound} Each pixel can be treated as an independent (for Poisson noise) experiment, and hence the Fisher information for the whole frame is given by
\begin{align}
    \text{FI}_{\text{Pois}} [\theta] &= \sum_j \frac{\left(\partial_\theta {I}_j \right)^2}{{I}_j}\\
     &= \norm{\frac{\partial_\theta I}{\sqrt{I}}}^2\\
    &= N_{\mathrm{ph}}\norm{\frac{2 \Re\left\{E_o^* \partial_\theta E_o\right\}}{|E_o|}}^2.
\end{align}
Observe that $\Re\left\{E_o^* \partial_\theta E_o\right\} = |E_o| |\partial_\theta E_o| \cos(\dots) \leq |E_o| |\partial_\theta E_o|$, and hence
\begin{align}
    \frac{\text{FI}_{\text{Pois}} [\theta]}{N_{\mathrm{ph}}} &\leq 4 \norm{|\partial_\theta E_o|^2} = 4,
\end{align}
where the last step follows the same argument as \cite{paterson2008towards}: $\partial_{a_k}E_o = i\mathcal{C}\{W_k E_i\}$, which is a unitary transformation of an orthonormal set.

\subsection{Sensitivity to read noise}
Next we consider a read noise only case in order to compare with the proof in \cite{chambouleyron2023modeling}, but note that the proof is identical for a pixel with both Poisson and read noise by simply changing the variance from $\sigma_R^2$ to $I_j + \sigma_R^2$.
The distribution sampled is for a single pixel (indexed by $j$), and is a normal distribution i.e. $X\sim \mathcal{N}(I_j, \sigma_R^2)$, where $\sigma_R$ is the read noise.
The Fisher Information with respect to a parameter is then
\begin{align}
    \text{FI}_\theta &= \mathbb{E}_x \left[\left(\frac{\partial}{\partial \theta} \log f (x;\theta) \right)^2 \right],
\end{align}
where $x$ is the integration variable from the normal distribution
\begin{align}
    f(x;\theta) &= \frac{1}{\sqrt{2\pi\sigma_R^2}} \exp\left(-\frac{(x-I_j(\theta))^2}{2\sigma_R^2}\right).
\end{align}
Applying the chain rule and factoring out terms that do not depend on $x$ we obtain
\begin{align}
    \text{FI}_\theta &=  \left(\frac{1}{{2\sigma_R^2}} \frac{\partial I_j(\theta)}{\partial \theta}  \right)^2 \mathbb{E}_x \left[\left(  \frac{\partial}{\partial (I_j(\theta))} \left(-(x-I_j(\theta))^2\right)  \right)^2 \right]\\
    &= \left(\frac{1}{{\sigma_R^2}} \frac{\partial I_j(\theta)}{\partial \theta}  \right)^2 \mathbb{E}_x \left[\left(x-I_j(\theta)  \right)^2 \right]\\
    &= \frac{1}{{\sigma_R^2}} \left( \frac{\partial I_j(\theta)}{\partial \theta}  \right)^2\\
    &= \frac{(\partial_\theta I_j(\theta))^2 }{\sigma_R^2} .
\end{align}
This is for an individual pixel $j$. Again one can sum this over all pixels
\begin{align}
    \text{FI}_\theta = \sum_j\frac{(\partial_\theta I_j(\theta))^2 }{\sigma_R^2}.
\end{align}

The result agrees with Eq. (24) of \cite{chambouleyron2023modeling} after noting that our intensity $I$ is not normalised to have a sum of unity over the image (and instead sums to $N_{\mathrm{ph}}$).

\textbf{Proof of the bound} Similar to before, the bound is now given by 
\begin{align}
    \text{FI}_\theta &= \frac{1}{\sigma_R^2}\norm{\partial_\theta I}^2\\
    &= \frac{N_{\mathrm{ph}}^2}{\sigma_R^2}\norm{2 \Re\left\{E_o^* \partial_\theta E_o\right\}}^2\\
    &= \frac{4 N_{\mathrm{ph}}^2}{\sigma_R^2}\norm{\Re\left\{E_o^* \partial_\theta E_o\right\}}^2\\
    &\leq \frac{4N_{\mathrm{ph}}^2}{\sigma_R^2}\norm{\,|E_o| \,|\partial_\theta E_o|\,}^2.
\end{align}
This is still bounded to a value of $4N_{\mathrm{ph}}^2/\sigma_R^2$. To see this, consider the expansion of the norm $\sum_j |E_{o,j}|^2 |\partial_\theta E_{o,j}|^2$, which, by the Cauchy-Schwarz inequality, is bounded by 
\begin{align}
    \sqrt{\sum_j |E_{o,j}|^2 \sum_j |\partial_\theta E_{o,j}|^2} \leq 1, 
\end{align}
where the first sum is less than or equal to unity by energy conservation (equality for a lossless system) and the second follows the same argument as in the previous subsection. Hence, the bound for the Fisher information for phase estimation in the read noise only case is 4 units read noise variance per radian per photon. 

\subsection{Wavefront sensors under both amplitude and phase errors} \label{ref:amp_and_phase_derivation}

\subsubsection{Classical bound}

Instead suppose the electric field at the input is given by
\begin{align}
    E_i &= A \exp\left(i \sum_k (a_k W_k) - \sum_k (b_k W_k)\right),
\end{align}
where the second summation represents (log-)amplitude aberrations that exist on the same basis as the phase aberrations. 

Applying \autoref{eq:diffI} with the general Fisher information formula derived earlier (under Poisson noise only), we consider the quantity of the sum of the Fisher information of phase and amplitude
\begin{align}
    \text{FI}_{a_k} &+ \text{FI}_{b_k} = \sum_j \frac{(\partial_{a_k}I_j)^2+(\partial_{b_k}I_j)^2}{I_j}  \\
    &=4N_{\mathrm{ph}}^2 \sum_j \frac{\Re\{E_o^*\mathcal{C}\left\{i W_k E_i \right\}\}^2 + \Re\{E_o^*\mathcal{C}\left\{- W_k E_i \right\}\}^2}{I_j}  \\
    &=4N_{\mathrm{ph}}^2 \sum_j \frac{\Im\{E_o^*\mathcal{C}\left\{W_k E_i \right\}\}^2 + \Re\{E_o^*\mathcal{C}\left\{- W_k E_i \right\}\}^2}{I_j}  \\
    &= 4N_{\mathrm{ph}} \sum_j \frac{|E_o^*\mathcal{C}\left\{W_k E_i \right\}|^2 }{|E_o|^2}  \\
    &= 4N_{\mathrm{ph}} \sum_j|\mathcal{C}\left\{W_k E_i \right\}|^2.
\end{align}
Since $W_k$ is orthonormal on the aperture, and $\mathcal{C}$ is unitary, the value of this integral is indeed $4N_{\mathrm{ph}}$, for all indices $k$. This indicates the total information over amplitude and phase is constant for an arbitrary, lossless wavefront sensor. This implies there is always a trade-off between amplitude and phase sensitivity.

\subsubsection{Quantum bound}
In the quantum case the field is 
\begin{align}
    \psi(x,y) &= A(x,y) \exp\left(i \sum_k (a_k W_k) - \sum_k (b_k W_k)\right).
\end{align}

The inner product in this space is defined as before. As in \cite{trzaska2026fundamental}, the quantum Fisher information matrix has elements $mn$ given by
\begin{align}
    \text{FI}\mathcal{^Q}_{mn} &= 4\Re\{\inner{\partial_m\psi}{\partial_n\psi} - \inner{\partial_m\psi}{\psi}\inner{\psi}{\partial_n\psi}\},
\end{align}
where $m$ and $n$ represent a phase or amplitude coefficient index and the inner product is as before but the first term now includes a complex conjugate (previously all terms were real).
First we compute the phase-phase term for the same index
\begin{multline}
\text{FI}\mathcal{^Q}_{a_ka_k} = 4\Re \bigg\{ \iint\left| \mathcal{C}\left\{W_k E_i \right\}\right|^2 \diff u \diff v \\
- \inner{\mathcal{C}\left\{i W_k E_i \right\}}{\mathcal{C}\left\{E_i \right\}}\inner{\mathcal{C}\left\{E_i \right\}}{\mathcal{C}\left\{i W_k E_i \right\}} \bigg\}.
\end{multline}
Similarly for the amplitude-amplitude term

\begin{multline}
\text{FI}\mathcal{^Q}_{b_kb_k} = 4\Re \bigg\{ \iint\left| \mathcal{C}\left\{W_k E_i \right\}\right|^2 \diff u \diff v \\
- \inner{\mathcal{C}\left\{- W_k E_i \right\}}{\mathcal{C}\left\{E_i \right\}}\inner{\mathcal{C}\left\{E_i \right\}}{\mathcal{C}\left\{- W_k E_i \right\}} \bigg\}.
\end{multline}

Notice that the second term in each integral is identical but with an opposite sign, as $i$ and $-1$ can be removed from the inner products. 

Following \cite{trzaska2026fundamental}, the information of each component is not simply summed, but rather weighted by a matrix 
\begin{align}
    C &= 
    \begin{bmatrix}
        I_{K}\cos^2(\theta) & 0 \\
        0 & I_{K}\sin^2(\theta)
    \end{bmatrix},
\end{align}
with $I_{K}$ as the identity matrix of size $K$ (the number of modes), and $\theta$ as a free parameter that determines how much weight is assigned to either phase or amplitude aberrations. As in the classic case, we assume equal weight to each ($\theta = \pi/4$). The weighted term then is
\begin{align}
    \frac{\text{FI}\mathcal{^Q}_{a_ka_k} + \text{FI}\mathcal{^Q}_{b_kb_k}}{2} &= 4,
\end{align}
where again we have used the orthonormality of $W_k$ and the unitary transformation properties of $\mathcal{C}$. This completes the proof that the trade-off between amplitude and phase information gain persists even when viewed through the lens of quantum information.

\subsection{Off-diagonal terms in the Fisher information matrix}\label{app:offdiag}
We can also explore the cases in which a system could have non-zero off-diagonal values in the Fisher information matrix, even when the phase basis is orthonormal. For brevity of notation, we switch to integration with infinitesimal pixels, where the integration is over the detector area. Considering the off-diagonal term, we obtain
\begin{align}
    \text{FI}_{a_ka_l} &= \int \frac{\partial_{a_k}I\partial_{a_l}I}{I} \diff x'\diff y'\\
    &= 4N_\mathrm{ph}\int  \frac{\Re\{E_o^*\mathcal{C}\{iE_iW_k\}\}\Re\{E_o^*\mathcal{C}\{iE_iW_l\}\}}{|E_o|^2}\diff x'\diff y'\\
    &= 4N_\mathrm{ph}\int \frac{\Im\{E_o^*\mathcal{C}\{W_k\}\}\Im\{E_o^*\mathcal{C}\{W_l\}\}}{|E_o|^2}\diff x'\diff y'\\
    &= 2N_\mathrm{ph}\int \frac{|E_o|^2\Re\{E_k^*E_l\} - \Re\{(E_o^*)^2\}E_kE_l}{|E_o|^2}\diff x'\diff y'\\
    &= 2N_\mathrm{ph} \int \Re\left\{E_kE_l^* - \frac{E_o^*E_kE_l}{E_o}\right\}\diff x'\diff y'\\
    &= 2N_\mathrm{ph} \Re\left\{ \inner{E_k}{E_l} - \inner{E_k}{E_l^*\frac{E_o}{E_o^*}}\right\},
\end{align}
where $E_k=\mathcal{C}\{W_k\}$. For a lossless system, recall the transformation $\mathcal{C}$ is unitary, and hence the first term is zero. Considering the second term, we note that $E_o/E_o^*=e^{2i\alpha_o}$ for an arbitrary $E_o=re^{i\alpha_o}$. 
One way to make the second term zero is to have
\begin{align}
    E_l^* e^{2i\alpha_o} &= E_l\\
    e^{i(-\alpha_l+2\alpha_o)} &= e^{i\alpha_l}\\
    \alpha_o=\alpha_l.
\end{align}
This is not possible for every $l$, since then the modes would not be orthogonal. If, however, both $\alpha_o$ is constant over the detector and $E_m$ is real for all $m$, the above constraint is not required and the second term is in fact zero everywhere.

From this we conclude that, when limited to a unitary/lossless system, the only wavefront sensor that has off-diagonal zero values in the Fisher information matrix is the \gls{PAWS}. The ideal wavefront sensor used in this work is a special case that maximises the on-diagonal values. It may be possible for lossy systems to have zero values in off-diagonal positions, but we do not explore this further as we are interested in systems that use all of the information.

\section{Additional results}
\subsection{Radial profiles of apodisations for different apertures}
\begin{figure}[h]
    \centering
    \includegraphics{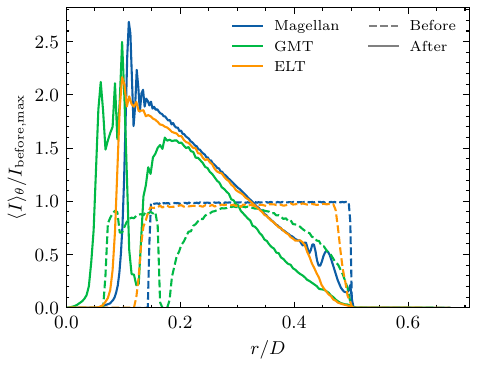}
    \caption{Radial profiles from the original aperture (before) and the apodised intensity profile (after, as shown in \autoref{fig:diff_apert}). }\label{fig:different_apertures_radial_profiles}
\end{figure}

\autoref{fig:different_apertures_radial_profiles} illustrates a radial profile (averaged over all angles) of the intensity for each aperture explored in \autoref{fig:diff_apert}. As expected, the intensity changes from a (nearly) top hat function to a decaying function. In the case of the Magellan aperture, some Fresnel propagation effects are noticeable at the edge.

\subsection{Stellar magnitude (normalised)}
\begin{figure}[h]
    \centering
    \includegraphics[scale=0.92]{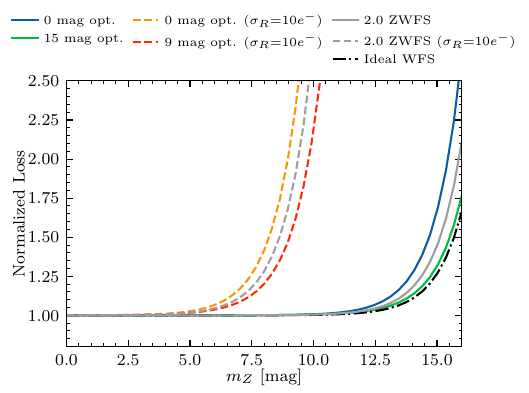}
    \caption{Same data as top of \autoref{fig:mag_change}, but normalised to the value at the bright limit for each sensor. The faint PIAA-ZWFS degrades slower than a ZWFS in both the low and high read noise cases, and closely follows the ideal WFS at the same read noise. }\label{fig:mag_change_normalised}
\end{figure}

To explore the rate of degradation at low photon count, we re-plot \autoref{fig:mag_change} in \autoref{fig:mag_change_normalised}, using the loss value normalised to each sensor's own performance at $m_Z = 0$.

\subsection{Wavefront reconstruction with open loop turbulence}
Here we present a version of \autoref{fig:recon_errors} with a $-11/3$ power law exponent, representing open loop turbulence. The result has the same behaviour between the different wavefront sensors but slightly different dynamic range.

\begin{figure}[h]
    \centering
    \includegraphics[scale=0.92]{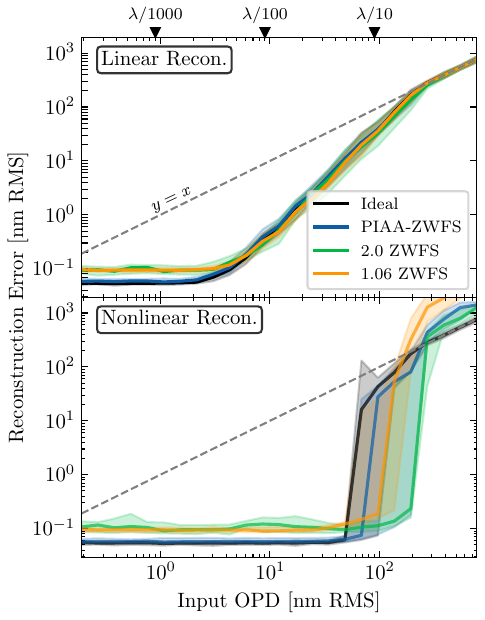}
    \caption{Same as \autoref{fig:recon_errors}, with a $-11/3$ power law exponent. Again, our proposed method performs much better than the classical \gls{ZWFS} when photon-noise-limited whilst maintaining a dynamic range comparable to the ideal wavefront sensor.}\label{fig:recon_errors_binned_113}
\end{figure}

\subsection{Generalisations of information bound saturation}\label{app:generalisation}
\begin{figure*}[h]
    \centering
    \includegraphics{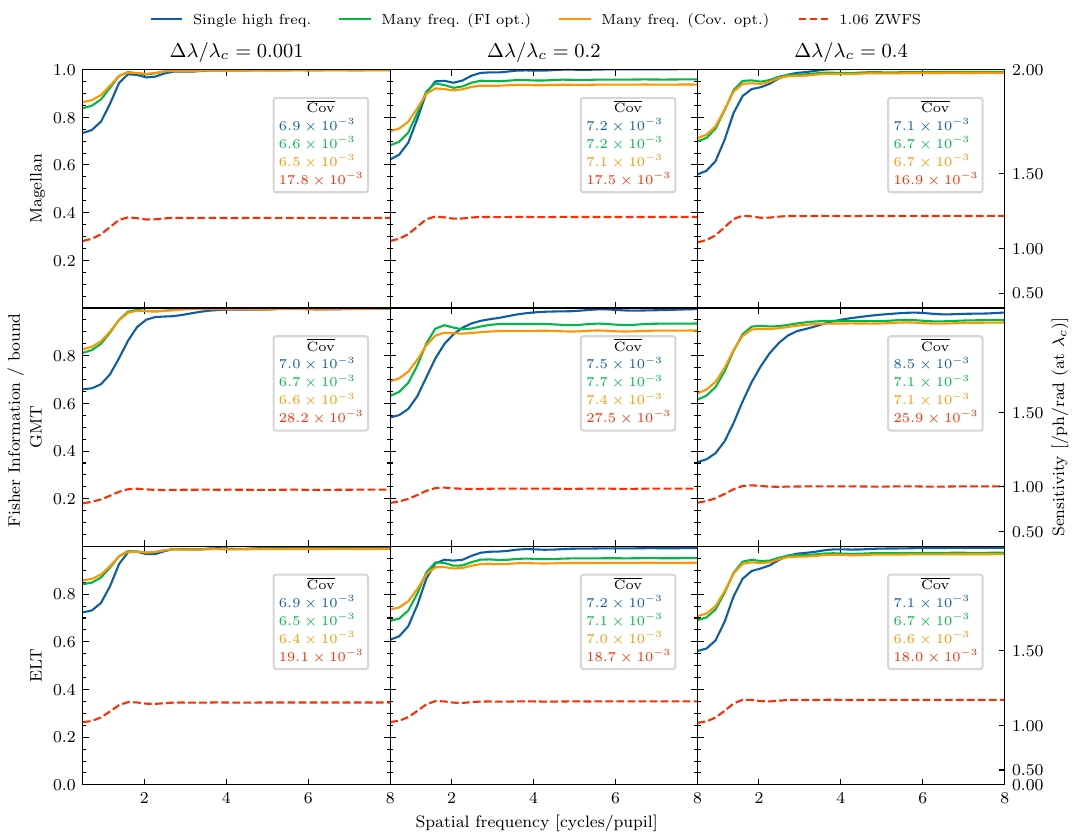}
    \caption{Same experiment as \autoref{fig:max_diff_freqs}, but for varying aperture and bandwidth. Values in the box show the average covariance of each design. The PIAA-ZWFS is generally able to saturate the information limit at high frequencies, but this is not optimal for the (equally weighted) covariance of the modes.} \label{fig:diff_metrics_grid}
\end{figure*}

\autoref{fig:diff_metrics_grid} shows repetitions of the experiment presented in \autoref{sec:infosat} for varying aperture and bandwidth (all done with a central wavelength $\lambda_c=1\,\mu$m). At larger bandwidths we found the loss space to have several local minima in covariance, so the ``FI opt.'' case was run first and used to hot-start the ``Cov. opt'' model. We observe the same trend as before -- namely that it is possible to maximise the high frequencies to virtually saturate the bound, but a more holistic solution finds a compromise between high and low frequencies. 

\end{appendix}
\end{document}